\documentstyle{laa}
\newhelp\stablestylehelp{You must choose a style between 0 and 3.}%
\newhelp\stablelinehelp{You should not use special hrules when stretching
a table.}%
\newhelp\stablesmultiplehelp{You have tried to place an S-Table inside another
S-Table.  I would recommend not going on.}%
\newdimen\stablesthinline
\newdimen\stablesthickline
\newif\ifstablesborderthin
\stablesborderthinfalse
\newif\ifstablesinternalthin
\stablesinternalthintrue
\newif\ifstablesomit
\newif\ifstablemode
\newif\ifstablesright
\stablesrightfalse
\newdimen\stablesbaselineskip
\newdimen\stableslineskip
\newdimen\stableslineskiplimit
\newcount\stablesmode
\newcount\stableslines
\newcount\stablestemp
\newcount\stablescount
\newcount\stableslinet
\newcount\stablestyle
\def\stablesleft{\quad\hfil}%
\def\stablesright{\hfil\quad}%
\catcode`\|=\active%
\newcount\stablestrutsize
\newbox\stablestrutbox
\setbox\stablestrutbox=\hbox{\vrule height10pt depth5pt width0pt}
\def\stablestrut{\relax\ifmmode%
                         \copy\stablestrutbox%
                       \else%
                         \unhcopy\stablestrutbox%
                       \fi}%
\newdimen\stablesborderwidth
\newdimen\stablesinternalwidth
\newdimen\stablesdummy
\newcount\stablesdummyc
\newif\ifstablesin
\stablesinfalse
\def\begintable{\stablestart%
  \stablemodetrue%
  \stablesadj%
  \halign%
  \stablesdef}%
\def\stablesadj{%
  \ifcase\stablestyle%
    \hbox to \hsize\bgroup\hss\vbox\bgroup%
  \or%
    \hbox to \hsize\bgroup\vbox\bgroup%
  \or%
    \hbox to \hsize\bgroup\hss\vbox\bgroup%
  \or%
    \hbox\bgroup\vbox\bgroup%
  \else%
    \errhelp=\stablestylehelp%
    \errmessage{Invalid style selected, using default}%
    \hbox to \hsize\bgroup\hss\vbox\bgroup%
  \fi}%
\def\stablesend{\egroup%
  \ifcase\stablestyle%
    \hss\egroup%
  \or%
    \hss\egroup%
  \or%
    \egroup%
  \or%
    \egroup%
  \else%
    \hss\egroup%
  \fi}%
\def\stablestart{%
  \ifstablesin%
    \errhelp=\stablesmultiplehelp%
    \errmessage{An S-Table cannot be placed within an S-Table!}%
  \fi
  \global\stablesintrue%
  \global\advance\stablescount by 1%
  \message{<S-Tables Generating Table \number\stablescount}%
  \begingroup%
  \stablestrutsize=\ht\stablestrutbox%
  \advance\stablestrutsize by \dp\stablestrutbox%
  \ifstablesborderthin%
    \stablesborderwidth=\stablesthinline%
  \else%
    \stablesborderwidth=\stablesthickline%
  \fi%
  \ifstablesinternalthin%
    \stablesinternalwidth=\stablesthinline%
  \else%
    \stablesinternalwidth=\stablesthickline%
  \fi%
  \tabskip=0pt%
  \stablesbaselineskip=\baselineskip%
  \stableslineskip=\lineskip%
  \stableslineskiplimit=\lineskiplimit%
  \offinterlineskip%
  \def\borderrule{\vrule width \stablesborderwidth}%
  \def\internalrule{\vrule width \stablesinternalwidth}%
  \def\thinline{\noalign{\hrule height \stablesthinline}}%
  \def\thickline{\noalign{\hrule height \stablesthickline}}%
  \def\trule{\omit\leaders\hrule height \stablesthinline\hfill}%
  \def\ttrule{\omit\leaders\hrule height \stablesthickline\hfill}%
  \def\tttrule##1{\omit\leaders\hrule height ##1\hfill}%
  \def\stablesel{&\omit\global\stablesmode=0%
    \global\advance\stableslines by 1\borderrule\hfil\cr}%
  \def\el{\stablesel&}%
  \def\elt{\stablesel\thinline&}%
  \def\eltt{\stablesel\thickline&}%
  \def\elttt##1{\stablesel\noalign{\hrule height ##1}&}%
  \def\elspec{&\omit\hfil\borderrule\cr\omit\borderrule&%
              \ifstablemode%
              \else%
                \errhelp=\stablelinehelp%
                \errmessage{Special ruling will not display properly}%
              \fi}%
  \def\stmultispan##1{\mscount=##1 \loop\ifnum\mscount>3 \stspan\repeat}%
  \def\stspan{\span\omit \advance\mscount by -1}%
  \def\multicolumn##1{\omit\multiply\stablestemp by ##1%
     \stmultispan{\stablestemp}%
     \advance\stablesmode by ##1%
     \advance\stablesmode by -1%
     \stablestemp=3}%
  \def\multirow##1{\stablesdummyc=##1\parindent=0pt\setbox0\hbox\bgroup%
    \aftergroup\emultirow\let\temp=}
  \def\emultirow{\setbox1\vbox to\stablesdummyc\stablestrutsize%
    {\hsize\wd0\vfil\box0\vfil}%
    \ht1=\ht\stablestrutbox%
    \dp1=\dp\stablestrutbox%
    \box1}%
  \def\stpar##1{\vtop\bgroup\hsize ##1%
     \baselineskip=\stablesbaselineskip%
     \lineskip=\stableslineskip%
     \lineskiplimit=\stableslineskiplimit\bgroup\aftergroup\estpar\let\temp=}%
  \def\estpar{\vskip 6pt\egroup}%
  \def\stparrow##1##2{\stablesdummy=##2%
     \setbox0=\vtop to ##1\stablestrutsize\bgroup%
     \hsize\stablesdummy%
     \baselineskip=\stablesbaselineskip%
     \lineskip=\stableslineskip%
     \lineskiplimit=\stableslineskiplimit%
     \bgroup\vfil\aftergroup\estparrow%
     \let\temp=}%
  \def\estparrow{\vfil\egroup%
     \ht0=\ht\stablestrutbox%
     \dp0=\dp\stablestrutbox%
     \wd0=\stablesdummy%
     \box0}%
  \def|{\global\advance\stablesmode by 1&&&}%
  \def\|{\global\advance\stablesmode by 1&\omit\vrule width 0pt%
         \hfil&&}%
  \def\vt{\global\advance\stablesmode by 1&\omit\vrule width \stablesthinline%
          \hfil&&}%
  \def\vtt{\global\advance\stablesmode by 1&\omit\vrule width
\stablesthickline%
          \hfil&&}%
  \def\vttt##1{\global\advance\stablesmode by 1&\omit\vrule width ##1%
          \hfil&&}%
  \def\vtr{\global\advance\stablesmode by 1&\omit\hfil\vrule width%
           \stablesthinline&&}%
  \def\vttr{\global\advance\stablesmode by 1&\omit\hfil\vrule width%
            \stablesthickline&&}%
  \def\vtttr##1{\global\advance\stablesmode by 1&\omit\hfil\vrule width ##1&&}%
  \stableslines=0%
  \stablesomitfalse}
\def\stablesdef{\bgroup\stablestrut\borderrule##\tabskip=0pt plus 1fil%
  &\stablesleft##\stablesright%
  &##\ifstablesright\hfill\fi\internalrule\ifstablesright\else\hfill\fi%
  \tabskip 0pt&&##\hfil\tabskip=0pt plus 1fil%
  &\stablesleft##\stablesright%
  &##\ifstablesright\hfill\fi\internalrule\ifstablesright\else\hfill\fi%
  \tabskip=0pt\cr%
  \ifstablesborderthin%
    \thinline%
  \else%
    \thickline%
  \fi&%
}%
\def\endtable{\advance\stableslines by 1\advance\stablesmode by 1%
   \message{- Rows: \number\stableslines, Columns:  \number\stablesmode>}%
   \stablesel%
   \ifstablesborderthin%
     \thinline%
   \else%
     \thickline%
   \fi%
   \egroup\stablesend%
\endgroup%
\global\stablesinfalse}
\def\ltsima{$\; \buildrel < \over \sim \;$}
\def\lsim{\lower.5ex\hbox{\ltsima}}
\def\gtsima{$\; \buildrel > \over \sim \;$}
\def\gsim{\lower.5ex\hbox{\gtsima}}
\def\mdot {\dot M}
\def\ls {LS~I~$+61\deg~303$}
\def\deg {^\circ}
\def\ergs {~erg$\,$s$^{-1}$}
\def\cms {~cm$\,$s$^{-1}$}
\def\cmdue {~cm$^{-2}$}
\def\cmtre {~cm$^{-3}$}
\def\gs {~g$\,$s$^{-1}$}

\def\msole{~M_{\odot}}

\def\aa #1 #2 {A\&A, {#1}, #2}
\def\aas #1 #2 {A\&AS, {#1}, #2}
\def\araa #1 #2 {ARA\&A, {#1}, #2}
\def\mon #1 #2 {MNRAS, {#1}, #2}
\def\apj #1 #2 {ApJ, {#1}, #2}
\def\apjs #1 #2 {ApJS, {#1}, #2}
\def\apjl #1 #2 {ApJ, {#1}, #2}
\def\aj #1 #2 {AJ, {#1}, #2}
\def\nat #1 #2 {Nature, {#1}, #2}
\def\pasj #1 #2 {PASJ, {#1}, #2}
\def\pasp #1 #2 {PASP, {#1}, #2}

\title{Radio pulsar and accretion regimes of rapidly rotating magnetic
neutron stars in early-type eccentric binaries}
\author{ S.~Campana\inst{1,2}
\and L.~Stella\inst{1,2}
\and S.~Mereghetti\inst{3}
\and M.~Colpi\inst{4}}

\begin{document}

\offprints{S.~Campana}

\institute{
{Osservatorio Astronomico di Brera, Via Brera 28, I-20121
Milano, Italy; \\ e-mail: (campana, stella)@astmim.mi.astro.it}
\and
{Affiliated to I.C.R.A.}
\and
{Istituto di Fisica Cosmica del C.N.R., Via Bassini 15, I-20133 Milano,
Italy; \\ e-mail: sandro@ifctr.mi.cnr.it}
\and
{Universit\`a degli Studi di Milano, Via Celoria 16, I-20133 Milano, Italy; \\
e-mail: colpi@astmiu.mi.astro.it}
}
\date{Received ; Accepted: 30 October 1994}

\maketitle
\label{sampout}

\begin{abstract}
Rapidly rotating magnetic neutron stars in eccentric binary systems
containing an early type star provide a unique opportunity to investigate
the interplay between radio pulsar, stellar wind and accretion phenomena.
We summarise the radio pulsar-dominated and the accretion-dominated
regimes, discussing how the transition from one regime to another can
take place as a result of the varying orbital distance and relative
velocity along the orbit, as well as changes of the wind characteristics.
We derive the conditions under which the two known B star/radio pulsar
binaries (PSR~1259--63 and PSR~J0045--7319) can undergo a transition to
the accreting regime. A strong increase of the mass loss ouflow from the
companion is required, just to cause the onset of accretion onto the
magnetospheric boundary.  We also show that the X--ray transient
A0538--66 is likely to undergo transitions from the accreting neutron
star regime, to the regime of accretion onto the magnetosphere.  These
two regimes might correspond to the high ($\gsim 10^{38}$\ergs) and the
low-luminosity ($< 10^{38}$\ergs) outbursts observed from this source. A
radio pulsar might become detectable in the long quiescent states of
A0538--66. A new model of the enigmatic high-energy binary \ls\ involving
accretion onto the magnetosphere is also presented.
\end{abstract}

\section{Introduction}

Two radio pulsars with massive B-star companions have been discovered in
the last few years: PSR~1259--63 and PSR~J0045--7319 (Johnston et al.
1992; Kaspi et al. 1994). Both are in highly eccentric orbits ($e>0.8$)
and likely represent the progenitors of high mass X--ray binaries
(Bhattacharya \& van den Heuvel 1991).  These systems are of great
importance for the investigation of the interplay between pulsar activity
and stellar wind, including possible transitions to the accretion regime
(e.g. Lipunov 1992; Kochanek 1993).  On the other hand, during the
quiescence intervals of X--ray transient binaries, accretion onto the
fast rotating neutron stars can stop completely and radio pulsar activity
may set in (e.g. Stella et al. 1994).

In Section 2 we briefly outline the different regimes of a radio pulsar
moving in the companion star's wind in an eccentric orbit.  Depending on
the neutron star spin and magnetic field, on the orbital parameters and
on the stellar wind characteristics,  the neutron star can alternate the
behaviour of a radio pulsar to that of an accreting X--ray source
(Illarionov \& Sunyaev 1975; Lipunov 1992).  These considerations are
applied to the two radio pulsars with B companions (Sections 3 and 4) and
to two peculiar X--ray binaries (A0538--66 and LS~I~$+61\deg~303$;
Sections 5 and 6). Our results are summarised in Section 7.

\section{Radio pulsar activity versus accretion}

Different regimes are possible for a radio pulsar immersed in the wind of
its companion star. If the radio pulsar is strong enough, a shock forms
in the interaction of the relativistic pulsar wind with the companion
star wind, well outside the accretion radius $r_{acc}=2\,G\,M/ v_{rel}^2$
($M$ is the neutron star mass and $v_{rel}$ the relative velocity between
the neutron star and the stellar wind matter). The condition for this
``pulsar radiation barrier" to work is obtained by equating the stellar
wind ram pressure and the pulsar radiation pressure at the accretion
radius:
$$
\mdot<{{f\,L^{sd}}\over {c\,v_{rel}}}\simeq 3.3\times 10^{17}\,f\,L^{sd}_{35}
\,v_7^{-1} {\rm \ g\,s^{-1}}$$
$$\simeq 1.3\times 10^{16}\,f\,\mu_{29}^2\,P_{-1}^{-4}
\,v_7^{-1} {\rm \ g\,s^{-1}} \ , \eqno(1)$$
\noindent where $\mdot$ is the mass capture rate, $L^{sd}_{35}$ the spin
down luminosity in units of $10^{35}$\ergs, $\mu_{29}$ the magnetic
\begin{table*}
\caption{Parameters of the neutron star binaries discussed in
the text.}
\smallskip
\stablesthinline=0pt
\stablesborderthintrue
\medskip
\begintable
Name \|$P$ \|$\mu$ \|$L^{sd}$ \| $\ e$ \| $P_{orb}$ \| $r_{per}$
\| \ $M_{*}$ \| $d$ \el
\| (ms) \| ($10^{29}\ {\rm G\,cm^{3}}$) \| ($10^{35}\ {\rm erg\,s^{-1}}$)\|
\| (d) \| (cm) \| ($\msole$) \| (kpc) \elt
PSR 1259--63    \| 47.8 \| 3.3 \| 8.3                \| 0.87 \| 1237
                         \| $2.1\times10^{13}$ \| $\sim 11$ \| 1.5 \el
PSR J0045--7319 \|926 \| 20.6 \| $2.2\times10^{-3}$ \| 0.81 \| 51.2
                         \| $3.0\times10^{12}$ \| $\sim 10$ \| 60 \el
A0538--66       \|69.2\| $3\times10^{-3}-1$ \| $< 0.17$
                                                     \|$>0.4\,$? \| 16.7
                         \|$ <3.0\times10^{12}\,$?  \| $\sim 12$ \| 55 \el
LS~I $+61^{\rm o} 303$ \| \|        \|               \|$>0.3\,$? \| 26.5
                         \|$ <4.3\times10^{12}\,$?  \| $\sim 10$ \| 2.3
\endtable
\end{table*}
\noindent dipole moment of the neutron star in units of $10^{29}{\rm \
G\,cm^{3}}$ and $P_{-1}$ the spin period in units of 0.1~s; $f$
represents the fraction of the pulsar pressure interacting with the
stellar wind (we assume $f=1$) and $v_7=v_{rel}/10^7$\cms. The mass
capture rate is related to the mass outflow from the companion star,
$\mdot_W$, by $\Omega\,r^2\,\mdot_W =\pi\,r_{acc}^2\,\mdot$ ($\Omega$ is
the angle subtended by the wind at the companion star and $r$ the
distance between the two stars).

The ``pulsar radiation barrier" can be overcome by the stellar wind
material if the mass capture rate $\mdot$ increases above the value in
Eq.(1), either as a consequence of the different orbital distance and
relative velocity, or as a result of substantial variations in the
stellar wind parameters. Inside the accretion radius the radial
dependence of the radio pulsar pressure ($\propto r^{-2}$) is less steep
than that of the pressure exerted by the stellar wind matter flowing
towards the neutron star ($\propto r^{-5/2}$ in the case of spherical
free fall or $\propto r^{-\alpha}$ with $51/20<\alpha<7/2$ in the case of
a standard accretion disk; Illarionov and Sunyaev 1975; Campana et al.
1995).
Therefore, if the ``pulsar radiation barrier" is won, the matter inflow
can proceed inside the light cylinder radius ($r_{\rm lc}={{c\,P}\over
{2\,\pi}}$, with $P$ the spin period), quenching the radio pulsar
emission (e.g. Illarionov \& Sunyaev 1975;  Lipunov 1992).  The motion of
the infalling matter becomes dominated by the rapidly increasing magnetic
field pressure ($\propto r^{-6}$) at the magnetospheric boundary
$$
r_{\rm m}\simeq 8.3\times 10^7\,\mu_{29}^{4/7}\,M_{1.4}^{-1/7}\,
\mdot_{17}^{-2/7} {\rm \ cm} \ , \eqno(2)
$$
\noindent where $M_{1.4}$ the neutron star mass in units of $1.4\,\msole$
and $\mdot_{17}$ the accretion rate in units of $10^{17}$\gs.

In general, accretion onto a rotating neutron star occurs only if the
centrifugal drag exerted by the magnetosphere on the accreting matter is
weaker than gravity (i.e. the ``centrifugal barrier" is open). If the
magnetosphere rotates at a super-Keplerian rate, matter cannot penetrate
the magnetospheric boundary and an accretion-luminosity of only
$$
L(r_{\rm m})=GM\mdot/r_{\rm m}\simeq 2.2\times 10^{35}\,\mu_{29}^{-4/7}
\,M_{1.4}^{8/7}\,\mdot_{17}^{9/7} {\rm \ erg\,s^{-1}} \eqno(3)
$$
\noindent is released (Stella et al. 1994; King \& Cominsky 1994).  In
this regime the fate of matter is uncertain: it can either accumulate
outside the magnetospheric boundary (possibly giving rise to a
quasi-steady atmosphere; Davies \& Pringle 1981) or be swang away by the
magnetospheric drag (as a result of either a supersonic or a subsonic
propeller; Davies \& Pringle 1981) at the expenses of the rotational
energy of the neutron star.  X--ray pulsations might be produced in this
regime as a result of the azimuthal asymmetry of the rotating
magnetospheric boundary. The ``centrifugal barrier" can be won only if
the accretion rate increases above
$$
\mdot\gsim 1.8\times 10^{18}\,\mu_{29}^2\,
M_{1.4}^{-5/3}\,P_{-1}^{-7/3}{\rm ~g\,s^{-1}}\ . \eqno(4)
$$
In this case accretion onto the neutron star surface takes place
releasing gravitational energy with a  much higher efficiency of
$L(R)=GM\mdot/R$, with $R$ the neutron star radius (we assume $R=10^6$\
cm). X--ray pulsations are likely to occur due to the channeling of the
accreting matter onto the neutron star magnetic poles.

When the ``centrifugal barrier" is closed, the magnetospheric boundary
expands for decreasing accretion rates. Eventually, $r_{\rm m}$ becomes
larger than the light cylinder radius and the radio pulsar mechanism can
resume. In the absence of accumulation of matter outside the
magnetosphere, this is expected to take place for
$$
L(r_{\rm lc})\simeq 8.5\times 10^{31}\,f\,\mu_{29}^2\,M_{1.4}^{1/2}\,
P_{-1}^{-9/2} {\rm \ erg\,s^{-1}} \ .\eqno(5)
$$
Due to its flatter radial dependence, the radio pulsar pressure sweeps
the material outside the accretion radius and the ``pulsar radiation
barrier" gives rise again to a shock front with the stellar wind. Note
that the luminosity in Eq.(3) is substantially higher than that in
Eq.(5), implying a higher luminosity threshold at the onset of accretion
and, therefore, a limit cycle behaviour.

When the radio pulsar mechanism resumes the rotational energy of the
neutron star must have decreased at least by the amount needed to eject
to infinity the matter accreted onto the magnetospheric boundary when the
``centrifugal barrier" was closed. This is equal to the total
gravitational energy released during the interval $\Delta t$ in which
accretion onto the magnetosphere takes place
$$
\Delta E=4\,\pi^2\,I\,{{\Delta P}\over{P^3}} \gsim \int_{\Delta t}
L(t)\,dt \ , \eqno(6)
$$
\noindent where $I$ is the moment of inertia of the neutron star and
$L(t)$ can vary between $L(r_{\rm m})$ and $L(r_{\rm lc})$. This
rotational energy loss can exceed the radio pulsar spin-down rate.

\section{PSR~1259--63}

PSR~1259--63 was the first radio pulsar to be found in a binary system
with an early type companion, the Be star SS2883 (Johnston et al. 1992,
1994;  see Table 1).  A campaign of multiwavelength observations was
carried out around the January 1994 periastron passage (the results for
the most part are still to be published).  The radio pulsar emission
disappeared during the two observed periastron passages (Johnston et al.
1992; Manchester et al. 1994), while it was clearly detected during the
rest of the orbit. These eclipses are due to an increase of the free-free
optical depth (Kochanek 1993; Lipunov et al. 1994) as testified by the
frequency dependence of the eclipse ingress (Manchester et al. 1994).
X--ray observations were carried out around apoastron (Cominsky, Roberts
\& Johnston 1994). X--rays were detected by ROSAT at orbital phases 0.6
and 0.73.  The average flux was about $40\%$ higher during the second
observation, which also showed a harder spectrum and evidence for
variability on timescales of days.  The conversion from observed count
rate to the unabsorbed X--ray luminosity is quite sensitive to the model
spectrum used to fit the data. A 0.1--2.4 keV luminosity of $\sim5\times
10^{32}\,d_{1.5}^2$\ergs\ is obtained in both observations for the
thermal bremsstrahlung and Raymond--Smith best fit models. On the other
hand, if the same data are fit with a power law spectrum, substantially
higher luminosities can be obtained (up to $\sim 8\times
10^{33}\,d_{1.5}^2$\ergs, in the case of the very soft spectrum of the
first observation). No pulsations at the radio period were detected.
GINGA obtained an upper limit corresponding to a 2--10 keV luminosity of
$6\times 10^{32}\,d_{1.5}^2$\ergs\ (for a Crab-like spectrum) at phase
0.31. The ROSAT observation at phase 0.45 gave a flux comparable to those
measured after periastron, at variance with earlier results reported by
Cominsky et al. (Belloni 1994, private communication).

In the case of PSR~1259--63, the ``pulsar radiation barrier" can be
overcome only for $\mdot\! > \!2.8\times 10^{18}\,v_7^{-1}$\gs\ [see
Eq.(1)], corresponding to an accretion-induced luminosity of $L(r_{\rm
m})\simeq 8.0\times10^{36}\,v_7^{-9/7}$\ergs\ [see Eq.(3)].  The
condition for the radio pulsar activity to resume and the inflowing
matter to be swept beyond the accretion radius (in the absence of
accumulation outside the magnetospheric boundary), corresponds to a
luminosity of $2.6\times 10^{34}$\ergs\ [see Eq.(5)].

King \& Cominsky (1994) proposed that the X--ray luminosity observed near
apoastron is produced by the release of gravitational energy down to the
magnetospheric boundary.  This possibility, however, presents serious problems.
Firstly, due to the radial dependence of the radio pulsar and the matter
pressure, the inflowing material cannot be stopped in a stable fashion just
outside the light cylinder radius.
If it is stopped inside the light cylinder, then the radio pulsar mechanism
is expected to be suppressed (e.g. Illarionov \& Sunyaev 1975; Lipunov 1992).
Secondly, the measured X--ray luminosities are
$\sim 50$ times lower than the
luminosity below which sweeping by the radio pulsar pressure sets in
(only for the X--ray observation around phase 0.6 this factor might
reduce to a value as low as $\sim 3$). Only if most of the energy is
released outside of the ROSAT and GINGA energy bands the observed
luminosities could be compatible with magnetospheric accretion.

The luminosities derived from the thermal model fits to the ROSAT data are
not incompatible with emission from the Be star. The corresponding value of
the X--ray to bolometric luminosity ratio, $L_X/L_{bol} \sim 3 \times
10^{-6}$, though higher than average, is well within the range of values
measured for B-type stars with or without emission lines (Meurs et al.
1992).
Alternatively, the observed X--ray emission might originate
in a discontinuity shock between the relativistic pulsar wind and the
matter outflowing from the Be companion. As shown by Tavani, Arons \& Kaspi
(1994), for PSR~1259--63 an X--ray luminosity of $\sim
10^{33}\,n_8^{1/2}\,v_7$\ergs\ can be obtained ($n_8$ is the density at the
shock in units of $10^{8}$\cmtre), with a power-law spectrum extending from
keV to MeV energies.

Accretion onto the magnetospheric boundary would take place if the
``pulsar radiation barrier" were overcome around periastron\footnote{For
accretion onto the neutron star surface to occur, a highly
super-Eddington accretion luminosity of $> 10^{40}$\ergs\ is needed.}.
This requires an accretion-induced luminosity, at least temporarily,
higher than $\sim10^{37}$\ergs, probably yielding a highly absorbed
X--ray spectrum ($N_H\gsim4\times10^{23}\,v_7^{-9/7} {\rm \ cm^{-2}}$).
This value implies also a very large free-free optical depth at radio
wavelengths  and the transition to the accretion regime would occur when
the radio pulsar is already undetectable.  The mass outflow from the Be
star can be derived from the required mass capture rate within the
accretion radius at periastron. For the equatorial disk mass loss, Waters
et al. (1988) estimate an outflow velocity law $v(r)\sim
v_0\,\bigl({r\over{R_*}}\bigr)$, with $v_0\simeq 5\times 10^5$\cms\ and
$R_*\simeq 7.6\times10^{11}$\ cm the stellar radius (taking a companion
mass of $M_*\simeq 11\msole$ and $R_*\propto M_*$). At large distances
from the Be star ($\sim 100\,R_*$) the equatorial wind achieves its
terminal velocity, $v_{\infty}$, of about three times the escape
velocity. Since at periastron $r=r_{per}\simeq 2.1\times 10^{13}$\ cm,
the pulsar velocity is comparable to the outflow wind velocity giving
$v_{rel}\simeq 2\times 10^7$\cms. Waters et al. (1987) estimate that the
characteristic solid angle of the Be equatorial wind is $\Omega\sim \pi$.
Therefore, we evaluate that a wind mass-loss rate of
$\mdot_W=1.3\times10^{-5} \msole {\rm\,yr^{-1}}$ is required to overcome
the ``pulsar radiation barrier".  This value is higher than the upper
limit of $\mdot_W <10^{-7} \msole {\rm\,yr^{-1}}$ on the mass outflow
rate obtained by Waters et al.  (1987) for $L_{bol}=2\times10^{38}$\ergs.
However, a substantially higher mass loss can occur during equatorial
shell-ejection episodes, such as those that cause the outbursts of
accreting X--ray pulsars with Be companions (e.g. A0535+26; Nagase et al.
1982). Further information can be derived from the reappearance of the
radio pulsar after periastron.  This occurred about one month after the
1990 periastron passage (Johnston et al. 1992). On the contrary if
accretion onto the magnetosphere sets in around periastron and $\mdot_W$
remains near or above the critical value, then the radio pulsar will
resume and start sweeping away the inflowing material not earlier than 8
months after periastron; moreover the free-free optical depth would be
$>1$ even at apoastron both for an isothermal and an adiabatic wind model
(Kochanek 1993 and references therein). Therefore the mass loss rate from
the Be star must decrease before the radio pulsar signal can be detected
again.

If shock emission is responsible for the X--ray emission at periastron a
luminosity of $\sim 10^{35}$\ergs\ is expected (for details see Tavani et
al. 1994). In any case the peak emission should not exceed the spin down
luminosity.

\section{PSR~J0045--7319}

This radio pulsar is the only known in the Small Magellanic Cloud.
Timing measurements show that it is in an eccentric orbit around a
massive B-type star showing, thus far, no evidence for emission lines
(Kaspi et al. 1994).  The lower spin-down luminosity and the smaller
distance to the companion (see Table 1) allow the transition from the
radio pulsar regime to the magnetospheric accretion regime to occur for a
less intense stellar wind than in PSR~1259--63. We consider the standard
spherical wind model for O and B stars, with a radial velocity law
$v(r)=v_{\infty} (1-R_*/r)^{1/2}$ (e.g. Castor \& Lamers 1979). At
periastron $v_{rel}\simeq 2\times10^8$\cms\ and, as long as the mass loss
rate from the B star is $\mdot_W\leq 1.4\times 10^{-7} \msole {\rm
\,yr^{-1}}$, the ``radiation pulsar barrier" cannot be overcome. No
regular eclipses of the radio pulses nor changes in the dispersion
measure were observed, implying that the wind from the B companion does
not exceed $10^{-11} \msole {\rm \,yr^{-1}}$ (note, however, that the
radio pulses were not detected on a few occasions; Kaspi et al. 1994).
This limit is somewhat lower than the mass loss expected for an isolated
B star (de Jager et al. 1988) and is strong enough to conclude that no
radio pulsar quenching can occur unless a very large variation of the
wind parameters, possibly a shell ejection, takes place.

\section{A0538--66}

The X--ray transient A0538--66 in the Large Magellanic Cloud contains the
accreting neutron star with the shortest known spin period (69 ms). The
periodic recurrence of most of its X--ray outbursts is highly suggestive
of a $16.7$~d  eccentric orbit around the Be star companion (White \&
Carpenter 1978; Skinner et al.  1982).  The high X--ray luminosity ($\sim
10^{39}$\ergs; Skinner et al. 1980) and the presence of pulsations
testify that during the bright outbursts the ``centrifugal barrier" is
open and accretion onto the neutron star surface takes place. As noted by
Skinner et al. (1982), the lower range of X--ray luminosities observed
during the bright outbursts implies an upper limit of $\mu_{29} \lsim 1$
[from Eq.(4)] (see also Maraschi, Traversini \& Treves 1983;  Stella,
White \& Rosner 1986).  On the other hand, if X--ray pulsations are
present also for the highest luminosities observed, then a small
magnetosphere is still present (i.e. $r_{\rm m} \!>\! R$); this implies
$\mu_{29} \gsim 3\times 10^{-3}$.  For A0538--66 the X--ray luminosity
resulting from  accretion onto the magnetosphere when the ``centrifugal
barrier" closes is $\sim 3 \times 10^{37}\,\mu_{29}^{2}$\ergs.

During the ROSAT all-sky survey two weak outbursts from A0538--66 were
detected, with average luminosities of $\sim 4$ and $\sim
2\times10^{37}$\ergs\ in the 0.1--2.4 keV range (Mavromatakis \& Haberl
1993). The emission was well fit by a black body spectrum with a
temperature $\sim 0.2$~keV, a characteristic radius $\lsim 5\times
10^7$~cm and a column density of $N_H\sim 10^{20}$\cmdue. This is unlike
the factor of $\sim 10$ more intense outbursts observed with Einstein,
that were characterised by harder and more absorbed spectra.  If
$\mu_{29} \sim 1$, the characteristic radius and X--ray luminosity
derived from the ROSAT data are close to the magnetospheric radius and
the maximum luminosity released by accretion onto the magnetosphere, when
the ``centrifugal barrier" is about to open.  In this case, values of
$N_H$ as low as those measured can be obtained for an accretion-disk flow
outside the magnetosphere.  Therefore, we suggest that the low-luminosity
and soft outbursts of A0538--66 could be powered by accretion onto the
neutron star magnetosphere, whereas during the outbursts with harder
spectra and $L\gsim 10^{38}$\ergs\ the ``centrifugal barrier" is open and
accretion can proceed down to the neutron star surface, generating
radiative energy and pulsations more efficiently.  An upper limit of
$\sim 5 \times 10^{34}$\ergs\ on the X--ray luminosity in quiescence has
been obtained with ROSAT. For this luminosity level, accretion onto the
neutron star surface is almost ruled out (it would require $\mu_{29}
\lsim 7\times 10^{-3}$), whereas accretion onto the magnetosphere can
take place over a wide range of allowed magnetic dipole moments. If the
accretion-induced luminosity decreases below
$4\times10^{32}\,\mu_{29}^2$\ergs, then the radio pulsar activity sets in
and sweeps away the inflowing matter beyond the accretion radius.

If in the extended quiescent state the Be equatorial disk does not
contribute significantly, the mass inflow towards the neutron star can be
calculated based only on the properties of the spherical wind, yielding a
rate of $\sim 3\times 10^{-9}\msole {\rm\,yr^{-1}}$ (Waters et al. 1988).
With $e\gsim0.4$ a value of $\mu_{29}\gsim 10^{-2}$ would be sufficient
for the radio pulsar to sweep away the incoming matter around apoastron.
We note that for $e = 0.4$ this limit on the magnetic dipole is also
sufficient to ensure that the ``pulsar radiation barrier" remains closed
even at periastron [see Eq.(1)] (a magnetic dipole 3 times larger would
be required for $e=0.7$). The free-free optical depth at 1~GHz is
expected to be $\ll 1$.  Therefore a 69~ms radio pulsar might be active
and detectable in the quiescent state of A0538--66.

\section{LS~I~+61$\deg$~303}

The Be star LS~I~$+61\deg~303$ shows strong radio outbursts with a
periodicity of 26.5~d, which are strongly suggestive of the presence of a
compact companion (Gregory \& Taylor 1978). Infrared observations
indicate a high mass loss in the stellar wind of the Be star ($1-4\times
10^{-7} \msole {\rm\,yr^{-1}}$; Waters et al. 1988). VLBI observations
show that the radio outbursts are produced by synchrotron emission from a
2 milliarcsecond double source expanding at $\sim 5\times10^7$\cms\
(Massi et al. 1994).  No periodic pulsations have been detected in the
radio emission, nor in its weak X--ray emission ($\sim
10^{33}\,d_{2.3}^2$\ergs\ in the 0.2--4 keV range with $d=2.3\,d_{2.3}$
kpc the distance; Bignami et al. 1981).  It has also been suggested that
this source is responsible for the $\gamma$--ray emission observed with
COS--B and EGRET from this region in the sky ($\sim
2.6\times10^{35}\,d_{2.3}^2$\ergs\ in the $>100$ MeV range;  Fichtel et
al. 1994).

Two models have been proposed: (a) shock emission from the relativistic
wind of a young neutron star colliding with the Be star wind (Maraschi \&
Treves 1981), similar to the model discussed in Section 3 for
PSR~1259--63; (b) super-Eddington accretion onto a neutron star close to
the periastron of an eccentric orbit (for the mass loss rate quoted above
this requires $e>0.3$; Taylor \& Gregory 1984; Taylor et al. 1992).

A problem with the latter model is that the measured X--ray and
$\gamma$--ray luminosities are orders of magnitude below the Eddington
limit. A possibility is that the accreting matter is stopped at the
magnetospheric radius, by the ``centrifugal barrier" and that a much
lower accretion luminosity is released.  In this case the matter ejected
through the centrifugal mechanism could explain the expansion of the
radio source.  By imposing that the measured X--ray luminosity is lower
than the luminosity for which the ``centrifugal barrier" opens [see
Eq.(3) and (4)], we obtain the condition $\mu_{29}^2 \,P_{-1}^{-3} \gsim
10^{-4}$ (or $\gsim 3\times 10^{-2}$ if the $\gamma$--ray luminosity is
used). The condition that the accreting material is not swept away by the
pulsar pressure [see Eq.(5)] gives instead $\mu_{29}^2\, P_{-1}^{-9/2}
\lsim 10$ (or $\lsim 3\times 10^3$, see above). We note that the high
mass inflow rate estimated by Taylor et al. (1992) for the periastron of
an $e=0.6$ orbit ($\mdot\simeq 2\times 10^{18}$\gs) is incompatible with
the magnetospheric accretion model, if the accretion-induced luminosity
is of the order of the observed X--ray luminosity. On the contrary the
observed $\gamma$--ray luminosity could be generated through this
mechanism in the case of a highly magnetic slowly rotating neutron star
($\mu_{29}\simeq 600$, $ 3\lsim P_{-1} \lsim 250$).

\section{Summary}

Due to the mass inflow variations resulting from the orbital motion
and/or the changing properties in the companion's wind, radio and X--ray
pulsars in eccentric early-type binaries offer the possibility of
studying the different regimes of a rapidly spinning neutron star
immersed in a stellar wind. Three different regimes are expected for
increasing mass inflow rates towards the neutron star: (i) radio pulsar;
(ii) accretion onto the neutron star magnetosphere; (iii) accretion onto
the neutron star surface. The transition {\it from} the radio pulsar
regime is expected to take place for mass inflow rates substantially
higher (a few orders of magnitude) than the transition {\it to} the radio
pulsar regime, therefore generating a characteristic limit cycle
behaviour. The transitions between the two accretion regimes do not
involve instead any limit cycle, such that, in principle, small
variations of the mass inflow rate can cause them to occur in both
directions.

The regime of magnetospheric accretion (i.e. down to $r_{\rm m}$) should
be characterised, in the optically thick regime, by temperatures in the
$10^{5}-10^{6}$ K range and radii comparable to the magnetospheric
radius. This regime might be  relevant to the quiescent emission of a
number of transient X--ray binaries containing a magnetic neutron star
(Stella et al. 1994;  Campana et al. 1995). We suggest that the
low-luminosity ($<10^{38}$\ergs) soft outbursts from the Be star/X--ray
pulsar transient A0538--66 are the result of magnetospheric accretion,
whereas the high-luminosity hard outbursts (during which 69~ms pulsation
are detected) are powered by accretion onto the neutron star surface.
This interpretation implies $\mu_{29}\sim 1$.  The occurence of
transitions between the two accretion regimes can be tested through the
monitoring of the temperature and radius variations during the rise or
the decay of very large outbursts.
The high energy emission from \ls\ might also arise from accretion onto
the magnetosphere of a neutron star.

Contrary to previous suggestions (King \& Cominsky 1994), we conclude
that the X--ray flux detected away from periastron in the eccentric radio
pulsar/Be star binary PSR~1259--63 is not powered by magnetospheric
accretion. More likely models include shock emission at the interface
between the pulsar and the Be star winds (Tavani et al. 1994) or emission
by an active Be companion.

No direct observational evidence has yet been found for a transition from
the radio pulsar regime to an accretion regime or viceversa. We have
shown that for the radio pulsars in PSR~1259--63 and PSR~J0045--7329 to
undergo a transition to the magnetospheric accretion regime a drastic
increase of the mass inflow rate must occur.  The radio pulsed emission
would resume only for much lower mass inflow rates, requiring in turn a
decrease of mass loss rate from the companion star (the variation caused
by the orbital motion alone would not be sufficient). The ejection of the
accreted material by the rotating magnetospheric boundary at the expenses
of the rotational energy of the neutron star is expected to cause a
pronounced spin-down during the magnetospheric accretion regime, which
can exceed the radio pulsar spin-down rate. In the absence of X--ray
pulsations in the magnetospheric accretion regime, this prediction can be
tested through radio pulse period measurements before and after a
magnetospheric accretion episode.

Concerning accreting neutron stars in X--ray binaries, a transition to
the radio pulsar regime is most likely to occur during the extended
quiescent states of X--ray transients (Stella et al. 1994; Campana et al.
1995). Among early-type systems, especially promising is the case of
A0538--66. In quiescence the inflow rate from the radial wind of the Be
star cannot prevent the transition to the radio pulsar regime to occur
around apoastron.  The resulting ``radio pulsar barrier" would then
inhibit accretion until substantially higher mass inflow rates are
achieved again. In particular if $\mu_{29} \sim 1$, the radio pulsar
signal should be detectable with current instrumentation.

\begin{acknowledgements}
We thank M. Salvati and M. Tavani for useful discussions.
SC gratefully acknowledges the receipt of an ASI fellowship.
This work was partially supported by ASI.
\end{acknowledgements}


\begin{thebibliography}{}

\bibitem[]{}
Bhattacharya, D., \& van den Heuvel, E.P.J. 1991, Phys. Rep., 203, 1

\bibitem[]{}
Bignami, G.F., Caraveo, P.A., Lamb, R.C., Markert, T.H. \& Paul, J.A.,
1981, \apj 247 L85

\bibitem[]{}
Campana, S., Colpi, M., Mereghetti, S., Stella, L. \& Tavani, M., 1995,
in preparation

\bibitem[]{}
Castor, J.I. \& Lamers, H.J.G.L.M., 1979, \apjs 39 481

\bibitem[]{}
Cominsky, L., Roberts, M. \& Johnston, S., 1994, \apj 427 978

\bibitem[]{}
Davies, R.E. \& Pringle, J.E., 1981, \mon 196 209

\bibitem[]{}
de Jager, C., Nieuwenhuijzen, H. \& van der Hucht, K.A., 1988, \aas 72 259

\bibitem[]{}
Fichtel, C.E. et al., 1994, ApJS in press

\bibitem[]{}
Gregory, P.C. \& Taylor, A.R., 1978, \nat 272 704

\bibitem[]{}
Illarionov, A.F., \& Sunyaev, R.A., 1975, \aa 39 185

\bibitem[]{}
Johnston, S. et al., 1992, \apj 387 L37

\bibitem[]{}
Johnston, S., Manchester, R.N., Lyne, A.G., Nicastro, L. \& Spyromilio,
J., 1994, \mon 268 430

\bibitem[]{}
Kaspi, V. et al., 1994, \apj 423 L43

\bibitem[]{}
King, A. \& Cominsky, L., 1994, ApJ in press

\bibitem[]{}
Kochanek, C.S., 1993, \apj 406 638

\bibitem[]{}
Lipunov, V.M., 1992, Astrophysics of neutron stars, Springer Verlag

\bibitem[]{}
Lipunov, V.M., Nazin, S.N., Osminkin, E.Yu. \& Prokhorov, M.E., 1994,
\aa 282 61

\bibitem[]{}
Manchester et al., 1994, {\it Millisecond pulsars: a decade of surprises},
eds. D.C. Backer, M. Tavani \& A.S. Fruchter, PASP in press

\bibitem[]{}
Maraschi, L., Traversini, R. \& Treves, 1983, \mon 204 1179

\bibitem[]{}
Maraschi, L. \& Treves, 1981, \mon 194 1P

\bibitem[]{}
Massi, M., Paredes, J.M., Estalella, R. \& Felli, M., 1993, \aa 269 249

\bibitem[]{}
Mavromatakis, F. \& Haberl, F., 1993 \aa 274 304

\bibitem[]{}
Meurs, E.J.A. et al., 1992, \aa 265 L41

\bibitem[]{}
Nagase, F. et al. 1982, \apj 263 814

\bibitem[]{}
Skinner, G.K. et al., 1980, \apj 240 619

\bibitem[]{}
Skinner, G.K. et al., 1982, \nat 297 568

\bibitem[]{}
Stella, L., Campana, S., Colpi, M., Mereghetti, S. \& Tavani, M., 1994,
\apj 423 L47

\bibitem[]{}
Stella, L., White, N.E. \& Rosner, R., 1986, \apj 308 669

\bibitem[]{}
Tavani, M., Arons, J. \& Kaspi, V., 1994, \apj 433 L37

\bibitem[]{}
Taylor, A.R. \& Gregory, P.C., 1984, \apj 283 273

\bibitem[]{}
Taylor, A.R., Kenny, H.T., Spencer, R.E. \& Tzioumis, A., 1992, \apj 395
268

\bibitem[]{}
Waters, L.B.F.M., Cot\`e, J. \& Lamers, H.J.G.L.M., 1987, \aa 185 200

\bibitem[]{}
Waters, L.B.F.M., Taylor, A.R., van den Heuvel, E.P.J., Habets, G.M.H.J.
\& Persi, P., 1988, \aa 198 200

\bibitem[]{}
White, N.E. \& Carpenter, G.F., \mon 183 11P

\end{thebibliography}
\end{document}